\documentclass[10pt,conference]{IEEEtran}
\IEEEoverridecommandlockouts
\usepackage{changepage}
\usepackage{framed}
\usepackage{tabularx}
\usepackage{subcaption}
\usepackage[acronym,nohypertypes={acronym}]{glossaries}
\makeatletter
 {\par\unskip\endMakeFramed}
\makeatother

\newenvironment{formal}{%
  \def\FrameCommand{%
    \hspace{1pt}%
    {\color{darkblue}\vrule width 2pt}%
    {\color{formalshade}\vrule width 4pt}%
    \colorbox{formalshade}%
  }%
  \MakeFramed{\advance\hsize-\width\FrameRestore}%
  \noindent\hspace{-1pt}% disable indenting first paragraph
  \begin{adjustwidth}{}{7pt}%
  \vspace{2pt}\vspace{2pt}%
}
{%
  \vspace{3pt}\end{adjustwidth}\endMakeFramed%
}

\newcounter{resultcounter}

\usepackage{hyperref,url}
\usepackage{xspace}
\usepackage{lipsum} 
\newcommand{\etal}[1]{#1~\textit{et al.}}

\newcommand{\citefig}[1]{Figure~\ref{fig:#1}}

\newcommand\GEAR{\texttt{GEAR}\xspace}

\usepackage{xcolor}
\usepackage{amssymb}
\usepackage{tikz}
\usepackage{pgfplots}
\usepackage{subcaption}
\pgfplotsset{compat=1.18}

\usepackage{flushend}
\newacronym{spl}{SPL}{{\emph{Software Product Line}}}
\newacronym{sple}{SPLE}{\emph{Software Product Line Engineering}}
\newacronym{uvl}{UVL}{\emph{Universal Variability Language}}
\newacronym{gen}{GenAI}{\textit{Generative AI}}
\newacronym{ai}{AI}{Artificial Intelligent}
\newacronym{mas}{MAS}{\textit{Multi-Agent System}}

\newacronym[
    \glslongpluralkey={\emph{Large Language Models}}, 
    \glsshortpluralkey={LLMs}
]{llm}{LLM}{\emph{Large Language Model}}
\newacronym[
    \glslongpluralkey={\emph{Software Product Lines}}, 
    \glsshortpluralkey={SPLs}
]{spls}{SPL}{\emph{Software Product Line}}
\newacronym[
    \glslongpluralkey={\emph{Feature Models}}, 
    \glsshortpluralkey={FMs}
]{fm}{FM}{\emph{Feature Model}}

\definecolor{gearfull}{HTML}{24619B}
\definecolor{gearpartial}{HTML}{8BC4E7}
\definecolor{gearnone}{HTML}{7D8A99}

\definecolor{gearnative}{HTML}{21866D}
\definecolor{gearsynth}{HTML}{7547A6}
\definecolor{gearadapt}{HTML}{E7B348}
\definecolor{gearunsupported}{HTML}{C64D72}

\definecolor{gearfull}{HTML}{2CA02C}
\definecolor{gearunverified}{HTML}{D62728}

\def\BibTeX{{\rm B\kern-.05em{\sc i\kern-.025em b}\kern-.08em
    T\kern-.1667em\lower.7ex\hbox{E}\kern-.125emX}}

\begin{document}

\title{GEAR: Feature-Based Configuration and Code Generation for Agentic Systems}

\author{
\IEEEauthorblockN{
Brell Sanwouo\IEEEauthorrefmark{1},
Nada Zine\IEEEauthorrefmark{1},
Paul Temple\IEEEauthorrefmark{2},\\
Clément Quinton\IEEEauthorrefmark{1}\IEEEauthorrefmark{3},
Romain Rouvoy\IEEEauthorrefmark{1}
}
\IEEEauthorblockA{
\IEEEauthorrefmark{1}Univ. Lille, CNRS, Inria, UMR 9189 CRIStAL, Lille, France\\
\IEEEauthorrefmark{2}Univ. Rennes, CNRS, Inria, IRISA, Rennes, France\\
\IEEEauthorrefmark{3}Institut Universitaire de France (IUF), Paris, France\\
\{brell.sanwouo,nada.zine,paul.temple,clement.quinton,romain.rouvoy\}@inria.fr
}
}

\maketitle

\begin{abstract}
Engineering \gls{mas} requires choosing agents, orchestration strategies, communication mechanisms, reasoning approaches, and memory configurations. These choices depend on framework-specific capabilities and abstractions, requiring developers to learn different APIs when exploring alternative implementations.
We present \GEAR, an open-source tool for configuring \glspl{mas} and generating framework-specific implementations from a common specification. 
\GEAR uses a feature-based representation to make configuration options and their constraints explicit, enabling users to validate specifications before code generation.
Through connectors for ten agentic frameworks, users generate projects from the same specification and can then inspect reports describing framework-specific adaptations.
A preliminary assessment of mapping feasibility identifies translation strategies for 271 of 320 evaluated feature–framework pairs (84.69\%). 
\GEAR supports developers prototyping agentic systems and researchers exploring \gls{mas} configurations across frameworks.
\texttt{GEAR} is available as open-source software at \url{https://github.com/brellsanwouo/gear-framework} and a video demo is available at \url{https://youtu.be/iwNi-oC2tCw}.
\end{abstract}


\glsresetall

\section{Introduction}
\label{sec:introduction}

Nowadays, developers and researchers increasingly build multi-agent applications using frameworks such as CrewAI~\cite{crewai2024}, Google ADK~\cite{googleadk2025}, and AutoGen~\cite{wu2024autogen}. 
%These frameworks expose many configuration choices related to agents, tools, memory, reasoning, and orchestration, often with dependencies and constraints that are distributed across APIs, code, configuration files, and documentation~\cite{wang2024survey,gao2025survey,guo2024large}.
%Moreover, although frameworks provide similar capabilities, they expose them through different abstractions, terminology, and APIs.
Although these frameworks provide similar capabilities (like definition of agents, memory, reasoning, or orchestration), they expose them through different abstractions, terminology, and APIs~\cite{wang2024survey,gao2025survey,guo2024large}.
%\pt{est-ce que la phrase précédente qui tente de merge avec les 2 phrases en commentaire est ok ?}
% \nz{perfect}
As a result, developers and researchers must both navigate a large configuration space and continuously learn framework-specific concepts and implementation choices making the design process time-consuming, error-prone, and difficult to reason about.

%Building these systems involves many configuration decisions concerning agents, orchestration workflows, communication protocols, reasoning strategies, and memory mechanisms~\cite{li2024survey}. However, these configuration options are often distributed across framework-specific APIs, code, configuration files, and documentation, and their dependencies and constraints remain implicit. As a result, developers often rely on iterative experimentation to construct valid configurations and explore alternative system designs, making the design process time-consuming, error-prone, and difficult to reason about.\pt{il faudrait une source/ref ici quand meme, les reviewers peuvent penser différemment...}
%The challenge becomes even bigger when moving across frameworks. Although agentic frameworks often provide similar concepts, they expose them through different abstractions, terminology, and APIs. A system designed for one framework, therefore, generally needs to be manually reinterpreted and reimplemented for another. This coupling limits portability and reuse, making it difficult to compare and reproduce \gls{mas} designs across frameworks.

%\nz{variability rapidement}
Software engineering has addressed similar problems in highly configurable systems through variability modeling~\cite{schaefer2012software}. In particular, \glspl{fm} provide explicit representations of configuration spaces and their constraints, together with automated reasoning capabilities for validating configurations~\cite{czarnecki2007feature,benavides2010automated}. However, to the best of our knowledge, no existing tool has leveraged \glspl{fm} to explicitly design and manage the variability of \glspl{mas}.

%\nz{on propose gear}
In this paper, we present \GEAR, an open-source tool that provides a common, framework-independent language for describing configurable \glspl{mas}. Users specify their agents, modules, workflows, and configuration choices once, and \GEAR validates this specification and generates executable implementations for selected target frameworks. 
%This allows developers to reuse the same \gls{mas} design across frameworks without having to learn and manually reimplement each framework-specific version. 
This allows developers to reuse the same \gls{mas} specification when exploring different target frameworks.
\GEAR connector-based architecture also separates the common specification from the implementation logic, enabling the integration of emerging frameworks and capabilities. A preliminary assessment across the $10$ supported frameworks identifies translation strategies for $271$ of $320$ evaluated feature--framework pairs ($84.69$\%).
%It also makes it easier to explore different frameworks, while its connector-based design allows \GEAR to evolve as new frameworks and capabilities emerge.

\textit{Envisioned users}: 
\texttt{GEAR} targets both developers building or prototyping LLM-based \glspl{mas} and researchers studying agentic system designs. Developers can use \texttt{GEAR} to explore configuration choices, detect invalid configurations before implementation, and generate executable implementations for different frameworks from the same specification. Researchers can use the common representation to systematically vary \gls{mas} designs, compare their implementations across frameworks, and reproduce experimental setups without manually reimplementing each system.

\section{Related Work}
\label{sec:background}

\textbf{Specification of \gls{mas}.}
Recent LLM-based multi-agent frameworks, such as CAMEL-AI~\cite{li2023camel}, AutoGen~\cite{wu2024autogen}, MetaGPT~\cite{hong2023metagpt}, CrewAI~\cite{crewai2024}, and Google ADK~\cite{googleadk2025}, provide abstractions for orchestrating agents through prompts, tools, and communication protocols. However, these frameworks primarily rely on code-level definitions, embedding system design decisions into implementation artifacts and offering limited support for reuse or systematic exploration of alternative configurations.
More recently, specification-oriented efforts, including Open Agent Specification~\cite{agentspec}, Agent Definition Language (ADL)~\cite{adl}, AgentSPEX~\cite{agentspex}, and H{\"a}rer~\cite{harer2025specification}, have introduced more declarative representations of agents and workflows. While these approaches contribute toward separating design from execution, they primarily describe individual system configurations or execution flows, with limited support for explicitly modeling variability across alternative \gls{mas} designs.

\textbf{Variability Modeling.}
%Variability modeling techniques have been applied across different domains to capture and manage variability in configurable software systems, including cloud computing~\cite{quinton2016saloon}, robotics~\cite{brugali2020software}, \emph{Internet of Things} (IoT)~\cite{trujillo2020multiple}, and fingerprinting~\cite{huyghe2025browserfm}.
%These techniques have also been extended to machine learning, including ML workflows, deep neural network architectures, and combined ML models~\cite{camillieri2016towards,ghofrani2019applying,ghamizi2019automated,gomez2024exploring}, and more recently to the configuration of \glspl{llm}~\cite{zine2026pimp}.
%\etal{Yoshimura} apply Software Product Line Engineering (SPLE) and feature modeling to industrial AI agents for root-cause analysis, capturing variability across equipment, models, data, and operational contexts~\cite{yoshimura2026managing}. Their work demonstrates the relevance of explicit variability management for configurable AI agents, but focuses on adapting a domain-specific agent across deployment contexts rather than representing and implementing \glspl{mas} across heterogeneous agentic frameworks.
%In contrast, \GEAR uses feature modeling to provide a common, framework-independent representation of \gls{mas} configurations and to generate executable implementations across multiple agentic frameworks.
Variability modeling has been widely used to capture and manage configuration choices in software systems, and has also been applied to machine learning systems~\cite{gomez2024exploring}. More recently, it has been also used to study the configuration of \glspl{llm}~\cite{zine2026pimp}. \etal{Yoshimura} study variability in industrial AI agents for root-cause analysis, illustrating the value of explicitly managing variations across equipment, models, data, and operational contexts~\cite{yoshimura2026managing}. While their work focuses on adapting a domain-specific agent to different operational contexts, \GEAR addresses variability at the \gls{mas} level through a common, framework-independent representation that supports configuration, validation, and generation across heterogeneous agentic frameworks.
\begin{figure}[h!]
    \centering
    \includegraphics[width=\linewidth]{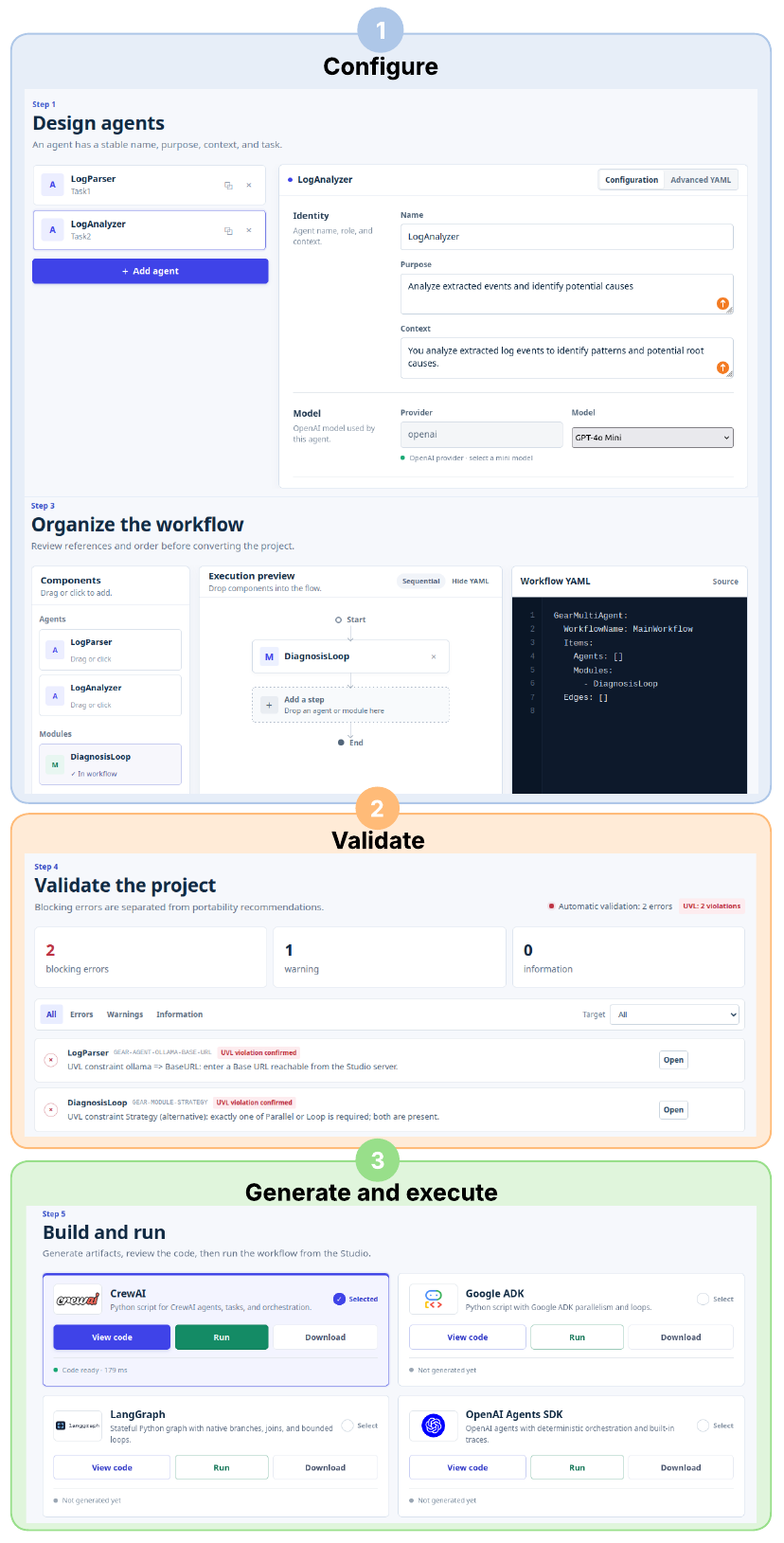}
    \caption{User workflow for the log-analysis demonstration in \GEAR Studio: (1) configure LogParser and LogAnalyzer and connect them sequentially; (2) validate the specification; (3) generate projects for CrewAI and Google ADK.}
    \label{fig:gear_features}
\end{figure}
%\nz{maj screen 2 avec exemple de validation parlant}
\section{The \GEAR Framework}
\label{sec:GEAR}

\subsection{\GEAR Core Features and Workflow}
\label{sec:workflow}

\GEAR provides a five-step workflow from \gls{mas} configuration to execution across different target frameworks, as shown in Fig.~\ref{fig:gear_features}. We illustrate this workflow with a log-analysis system comprising two agents: \textit{LogParser}, which parses logs, and \textit{LogAnalyzer}, which analyzes them to identify potential causes.

\textbf{Configure.}
Users define a \textit{framework-independent specification} through \GEAR Studio.
%or a YAML file.\pt{il y a un template de ce YAML file quand même ? Parce que... Sinon c'est un fichier crée from scratch et c'est pas très bon...} 
The specification comprises three main elements: (i) \textit{agents}, with properties such as their model, task, tools, memory, and execution controls; (ii) \textit{modules}, which define reusable orchestration structures such as parallel or iterative execution; and (iii) \textit{workflows}, which connect agents and modules and define their execution flow. In our example, the user configures the two agents and groups them into a \textit{DiagnosisLoop} module used in the workflow.

%\pt{ça se voit où sur l'interface de la figure 2?}
%\nz{il y'a un module qui regroupe les deux du coup on le voit pas, 'jai rajouté une explication dans le texte}
\textbf{Validate.}
%\GEAR checks the specification constraints, component references, and workflow consistency. For example, a reference to an undefined agent or an invalid combination of options is reported to the user. Invalid specifications must be fixed before framework selection. \cq{ex concret de contrainte (qu'on verrait dans la figure [en montrer moins du coup])?}
    \GEAR automatically validates agents, modules, and workflows, reporting errors such as references to undefined agents or incompatible configuration options. Users can then correct these errors before proceeding to framework selection. In our example, the user selects both parallel and loop execution for \textit{DiagnosisLoop}. Because these strategies are mutually exclusive, \GEAR reports the conflict and asks the user to select a single strategy.

%\bs{Durant la spécification, \GEAR valide automatiquement la configuration des agents, des modules et des workflows, en signalant les erreurs telles que les références à des agents non définis ou les options de configuration incompatibles. Les utilisateurs peuvent alors corriger ces erreurs avant de passer à la sélection du framework.  }

\textbf{Select Target Frameworks.}
%Users select one or several target frameworks, \eg CrewAI and Google ADK. We selected frameworks based on their availability, documentation, active maintenance, Python support, and ability to compose agents into workflows~\cite{wang2024survey,gao2025survey,guo2024large}. 
%For each target, \GEAR determines whether each selected property is directly identified, requires adaptation or is unresolved. For instance, a model configuration may map directly, while a loop stopping condition may require adaptation. Properties without a mapping are reported as unresolved and omitted from the generated implementation, and users must then inspect the conversion report to determine whether omitted properties are acceptable for their application.
Users select one of the ten target frameworks supported by \GEAR for each generation. The same specification can then be reused to generate implementations for other frameworks. In our example, the user first selects CrewAI and then repeats the generation with Google ADK, obtaining two implementations from the same log-analysis specification.

\textbf{Generate.}
\GEAR generates a separate executable project for each selected framework from the validated specification using templates. 
Each project contains source code, a configuration file and an execution entry point.
In our example, the same log-analysis system can be generated for both CrewAI and Google ADK using the same specification. %Each project includes a conversion report describing mappings, adaptations, and unresolved properties. \GEAR currently supports $10$ agentic frameworks.

\textbf{Inspect and Execute.}
Users can inspect the generated project to determine whether it meets their application requirements. 
They can then execute it through the command line or through \GEAR when execution is available.

\begin{figure*}[h!!]
    \centering
    \includegraphics[width=\textwidth]{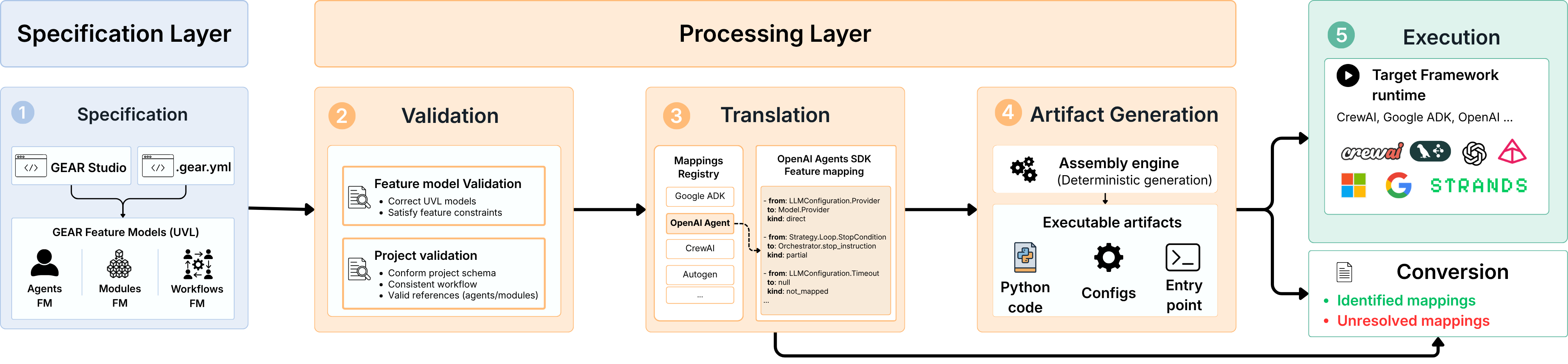}
    \caption{Architecture and processing pipeline of \GEAR.} % \cq{il reste supported unsupported à droite}}
    \label{fig:gear_architecture}
\end{figure*}

\begin{figure*}[h!]
    \centering
    \includegraphics[width=\textwidth]{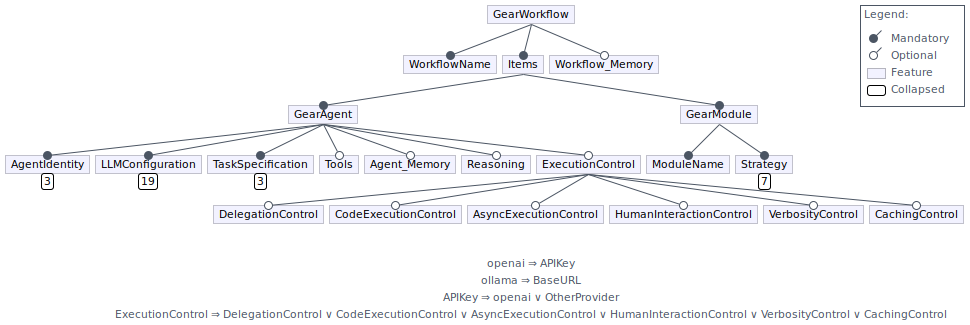}
    \caption{Excerpt of \GEAR's feature model.}
    \label{fig:gear_fm}
\end{figure*}

\subsection{\GEAR Architecture and Implementation}

Fig.~\ref{fig:gear_architecture} presents the architecture and processing pipeline supporting the workflow described above. \GEAR separates the common specification of a \gls{mas} from the mechanisms required to transform it into framework-specific implementations.

\textbf{Specification Layer.}
The specification layer (highlighted in blue in Fig.~\ref{fig:gear_architecture}) relies on feature modeling to represent the configurable properties of agents, modules, and workflows. A feature model organizes these properties hierarchically and specifies whether they are mandatory, optional, or alternative, together with dependencies and constraints governing their combinations~\cite{czarnecki2007feature}. Fig.~\ref{fig:gear_fm} illustrates an excerpt of \GEAR's common variability representation. In particular, $32$ features encode \gls{mas} concepts identified through a review of the literature and an analysis of ten agentic frameworks, selected based on their availability, documentation, active maintenance, Python support, and ability to compose agents into workflows~\cite{wang2024survey,gao2025survey,guo2024large}. 
In addition, the feature model covers agent identity, model and task configuration, tools, memory, reasoning, execution controls, and orchestration. 
Overall, it exhibits $53$ features and encodes approximately $2.4 \times 10^{8}$ valid configurations.
%The models are expressed in UVL~\cite{benavides2025uvl} and are independent of the target framework. 
% \pt{comment on peut savoir qu'elles ont été définies pour être indépendantes des frameworks ?}

% \pt{il faudra présenter le nombre de features booléennes, à valeur numérique, énumération, etc. dans le repo ou en annexe si il y a la possibilité et sous la forme d'un tableau je pense}

\textbf{Processing Layer.}
The processing layer (highlighted in brown in Fig.~\ref{fig:gear_architecture}) transforms this specification into framework-specific implementations. During the \textit{validation} phase, \GEAR checks the specification structure, feature-model constraints, and workflow consistency, and normalizes valid configurations into a common intermediate representation. During the \textit{translation} phase, a connector for each target framework applies explicit feature mappings and provides the manifest, templates, and assembly logic required by that framework. The \texttt{Assembly engine} then deterministically generates the executable artifacts, including source code, configuration files, and execution entry points. A conversion report records how selected features were mapped, adapted or unresolved, preserving traceability between the common specification and each generated implementation.

% \pt{cette notion de mapping me parait arrivée un peu trop abruptement... On va se demander à quoi ça correspond, qu'est-ce que ça veut dire partially supported et unresolved? Pourquoi il y a une validation alors, si il y a des choses qui peuvent être définies mais non-supportées ? Je pense qu'il serait peut-être plus simple de dire uniquement qu'il y a un rapport qui est généré pour le détail des étapes de génération des implémentations}

% \bs{parler d'une petite évaluation même légère -> voir GeMTest: A General Metamorphic Testing Framework}
% \bs{prévalidation -> taux de couverture ?}

\newcommand{\gearPropertySupportChart}[2]{%
    \begin{tikzpicture}
    \begin{axis}[
        xbar stacked,
        width=#1,
        height=#2,
        bar width=9pt,
        xmin=0, xmax=42,
        ymin=0.5, ymax=10.5,
        y dir=reverse,
        ytick={1,...,10},
        yticklabels={
            CrewAI,
            Google ADK,
            LangGraph,
            OpenAI Agents,
            Microsoft AF,
            Strands,
            PydanticAI,
            AutoGen,
            Semantic Kernel,
            Haystack
        },
        ytick style={draw=none},
        yticklabel style={font=\small},
        xtick={0,8,16,24,32},
        xlabel={Number of evaluated GEAR features},
        xlabel style={font=\small},
        xticklabel style={font=\small},
        xmajorgrids,
        grid style={black!12},
        axis y line*=left,
        axis x line*=bottom,
        legend columns=2,
        legend style={
            draw=none,
            at={(0.5,-0.26)},
            anchor=north,
            font=\small,
            column sep=4pt
        },
        area legend,
        clip=false,
        point meta=explicit symbolic,
        nodes near coords={\pgfplotspointmeta},
        nodes near coords align=center
    ]

    % Supported mappings
    \addplot[
        fill=gearfull,
        draw=none,
        every node near coord/.append style={
            font=\small,
            text=white
        }
    ] coordinates {
        (31,1) [31]
        (30,2) [30]
        (27,3) [27]
        (28,4) [28]
        (24,5) [24]
        (27,6) [27]
        (23,7) [23]
        (28,8) [28]
        (27,9) [27]
        (26,10) [26]
    };
    \addlegendentry{\(\checkmark\) Identified mapping}

    % Unresolved mapping
    \addplot[
        fill=gearunverified,
        draw=none,
        every node near coord/.append style={
            font=\small,
            text=white
        }
    ] coordinates {
        (1,1) [1]
        (2,2) [2]
        (5,3) [5]
        (4,4) [4]
        (8,5) [8]
        (5,6) [5]
        (9,7) [9]
        (4,8) [4]
        (5,9) [5]
        (6,10) [6]
    };
    \addlegendentry{\(\times\) Unresolved mapping}

    % Coverage percentages
    \node[anchor=west, font=\small] at (axis cs:33,1) {96.88\%};
    \node[anchor=west, font=\small] at (axis cs:33,2) {93.75\%};
    \node[anchor=west, font=\small] at (axis cs:33,3) {84.38\%};
    \node[anchor=west, font=\small] at (axis cs:33,4) {87.50\%};
    \node[anchor=west, font=\small] at (axis cs:33,5) {75.00\%};
    \node[anchor=west, font=\small] at (axis cs:33,6) {84.38\%};
    \node[anchor=west, font=\small] at (axis cs:33,7) {71.88\%};
    \node[anchor=west, font=\small] at (axis cs:33,8) {87.50\%};
    \node[anchor=west, font=\small] at (axis cs:33,9) {84.38\%};
    \node[anchor=west, font=\small] at (axis cs:33,10) {81.25\%};

    \end{axis}
    \end{tikzpicture}%
}

\begin{figure}[h!]
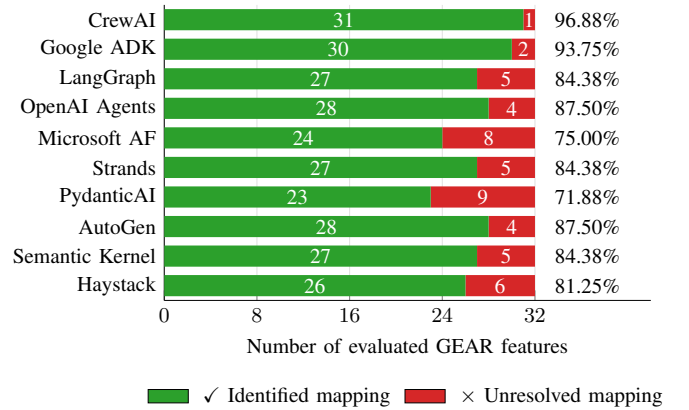

    \centering
    \resizebox{1\columnwidth}{!}{%
        \gearPropertySupportChart{\columnwidth}{6cm}%
    }
    %\caption{Coverage of identified mappings between \GEAR features and the target frameworks. The percentages indicate the proportion of features with an identified mapping for each framework.}
    \caption{Coverage of identified mappings between \GEAR features and target frameworks (\% of features mapped per framework). }
    \label{fig:gear-property-support}
\end{figure}

%\section{Validation Study}
\section{Preliminary Mapping Assessment}
\label{sec:validation}

We conducted a preliminary mapping assessment across the ten frameworks, listed in \citefig{gear-property-support}. 
For each of the 32 evaluated \GEAR features, we examined whether a framework-specific implementation was feasible, yielding 320 feature–framework pairs. 
%\pt{on ne sait pas comment elles sont définies}, we examined whether a framework-specific implementation was feasible, yielding 320 feature–framework pairs. 
We considered a mapping as identified when we could determine how the feature could be implemented in the target framework, using a native mechanism, an adaptation, or synthesized code. 
Otherwise, we marked the mapping as unresolved, particularly when the feature was not exposed by the framework or remained implicit in its internal mechanisms. 
The identified mappings were then used to define and extend the framework-specific connectors of \GEAR.

Figure~\ref{fig:gear-property-support} presents the mapping results for 320 feature--framework pairs, obtained by evaluating each of the 32 \GEAR features against the ten selected frameworks. Overall, mappings were identified for 271 pairs (84.69\%), while 49 (15.31\%) remained unresolved. CrewAI achieved the highest coverage, with mappings identified for 31 \GEAR features (96.88\%) and one unresolved, followed by Google ADK with 30 (93.75\%) and two, respectively. At the other end, Microsoft Agent Framework had 24 identified mappings (75.00\%) and eight unresolved, while PydanticAI had 23 (71.88\%) and nine. A given \GEAR feature may be mapped to some frameworks while remaining unresolved in others.

% Figure~\ref{fig:gear-property-support} summarizes the mappings identified between the 32 features of \GEAR and the ten selected frameworks. Among the 320 feature--framework pairs, a mapping was identified for 271 (84.69\%), while unresolved mapping was found for the remaining 49 (15.31\%). CrewAI has the highest proportion of identified mappings (96.88\%), followed by Google ADK (93.75\%). In contrast, PydanticAI (71.88\%) and Microsoft Agent Framework (75.00\%) have the lowest proportions. These results show that most \GEAR features can be mapped to the selected frameworks, although an identified mapping does not necessarily guarantee full semantic preservation. \pt{j'ai toujours un problème avec ça... Ca m'irait si le message était de dire que l'on n'a pas pu intégrer 49 features dans GEAR par manque de temps... Mais là, le message que je lis c'est qu'il y a 49 features que l'on peut définir dans les fichiers YAML mais qui en fait ne peuvent pas se traduire dans des configs... C'est vraiment ça ou j'ai mal compris ?}

%\cq{As future work, we plan to conduct a user study in which participants will use \GEAR to specify and generate \gls{mas}. This study will assess the tool's usability, the effort required to complete these tasks, and the difficulties encountered by participants. MOVE TO NEXT SECTION ?}
To further assess \GEAR, we are conducting an ongoing user study in which participants use the tool to specify and generate \glspl{mas}. We evaluate usability, effort, and the difficulties they encounter.
\section{Limitations \& Future Work}
\label{sec:limitations}

While \GEAR demonstrates its ability to facilitate the configuration, validation, translation, generation, and execution of \gls{mas} across different frameworks, several limitations remain. First, the current implementation focuses on the core abstractions and functionalities shared by the considered frameworks. More advanced capabilities, such as multimodal inputs and streaming are currently out of scope. Many of these capabilities are highly framework-specific and were intentionally excluded to preserve a common abstraction level and enable a controlled comparison across frameworks.

Second, support for a new target framework currently requires manually defining the corresponding connectors and mappings between \GEAR's framework-independent specification and framework-specific concepts. Although these mappings are reusable once implemented, their initial development requires both engineering effort and knowledge of the target framework. Reducing this effort, for instance through partially automated connector generation or mapping inference, is therefore an important direction for future work.

%Finally, \GEAR primarily guarantees consistency at the specification and translation levels. It does not attempt to eliminate semantic, runtime, or performance differences that may arise from the underlying frameworks, their execution models, or their implementations. 
\GEAR validates specifications against encoded constraints and reports target-specific translation limitations. Unresolved properties are omitted, so generated projects may differ from the requested specification.
Future work should therefore investigate these behavioral differences more systematically, extend the range of supported features, and evaluate \GEAR on larger and more representative real-world multi-agent applications.

\section{Conclusion}
\label{sec:conclusion}
In this paper, we presented \GEAR, an open-source tool for configuring and generating \glspl{mas} across heterogeneous agentic frameworks. By separating the specification of a \gls{mas} from its framework-specific implementation, \GEAR allows users to reuse the same \gls{mas} specification across different frameworks through framework-specific connectors. It also makes configuration choices explicit through its feature-based model and makes the support of new frameworks easier through its connector-based design. Overall, \GEAR contributes to bridging variability modeling with \gls{mas}, and opens new directions for their study and evaluation.
%\pt{je pense que la conclusion ne reflète pas assez le fait que ça aiderait les utilisateurs à réutiliser sans effort et de manière automatique les descriptions de leur MAS; que l'outil est extensible. La deuxième phrase surtout doit pouvoir être plus percutante en mettant en avant le fait que les SPL sont là vraiment pour simplifier le passage d'un framework à un autre sans effort supplémentaire.}
\textbf{Data Availability}
\label{data}
\texttt{GEAR} is open-source and available at \url{https://github.com/brellsanwouo/gear-framework}. 
A demonstration video presenting an end-to-end case study is also available at \url{https://youtu.be/iwNi-oC2tCw}. 

%\section*{Acknowledgment}

\bibliographystyle{IEEEtran}
\bibliography{bibfile.bib}

\end{document}